\documentclass[preprint]{aastex}

\shorttitle{Cosmic Ray Iron Flux}
\shortauthors{Clem et al.}
\begin{document}


\title{A Measurement of the Flux of Cosmic Ray Iron at 5$\times$10$^{13}$ eV}


\author{J.~Clem\altaffilmark{1}, W.~Droege\altaffilmark{1,2}, P.A. 
Evenson \altaffilmark{1,3}, H.~Fischer\altaffilmark{2}, G.~Green\altaffilmark{2},
D.~Huber\altaffilmark{1}, H.~Kunow\altaffilmark{2}, and 
D.~Seckel\altaffilmark{1}}

\altaffiltext{1}{Bartol Research Institute, University of Delaware, 
Newark, DE 19716}
\altaffiltext{2}{Christian-Albrechts-Universit\"at Kiel, D-24089 Kiel, Germany}
\altaffiltext{3}{National Science Foundation, Arlington, VA 22230, USA}
\email{droege@bartol.udel.edu}


\begin{abstract}
We present results from the initial flight of our Balloon Air
CHerenkov (BACH) payload. BACH detects air Cherenkov radiation
from cosmic ray nuclei as coincident flashes in two optical
modules. The flight (dubbed PDQ~BACH) took place on April
22, 1998 from Ft. Sumner, New Mexico. During an exposure of 2.75
hours, with a typical threshold energy for iron nuclei of
2.2$\times$10$^{13}$ eV, we observed several events cleanly
identifiable as iron group nuclei. Analysis of the data yields a
new flux measurement that is fully consistent with that reported
by other investigations. 
\end{abstract}




\section{Introduction} 

It is important to extend direct measurements of cosmic ray composition 
towards the knee of the cosmic ray spectrum. Determining composition in 
this region is one of the keys to understanding cosmic ray acceleration 
and propagation. Air shower techniques are a potential tool for measuring 
composition at the knee, but their interpretation ultimately relies on 
extrapolation of particle interaction models to regimes not directly 
tested in laboratory experiments. Emulsion techniques and electronic 
detectors must be improved before they can provide an independent 
calibration of air shower experiments, and so the potential for 
systematic error exists. We have designed an experiment (Evenson and 
Seckel 1995; Seckel et al. 1999) to determine the flux of iron nuclei at 
energies between 3.0$\times 10^{13}$ and 3.0$\times 10^{14}$ eV by 
detecting atmospheric Cherenkov radiation from the incoming nucleus 
before it undergoes a nuclear interaction. By achieving an absolute 
measurement of the iron flux, relatively free of the uncertainty in 
particle interaction models, the Balloon Air CHerenkov (BACH) technique 
can provide an independent calibration for air shower experiments, thus 
opening a path for composition measurements for energies at and above 
the knee. This technique was discussed by Sitte (1965) and Gough (1975), 
and used initially by Sood (1983; Sood and Panettieri, 1981). In this 
paper we report the result obtained from the initial flight of our 
developmental balloon payload (PDQ~BACH). 

\section{Technique}

A high energy iron nucleus entering the atmosphere has an interaction 
length of 13.2~gm/cm$^2$ (Hufner 1985). Before interacting, a 
sufficiently high energy nucleus will begin to radiate photons via the 
Cherenkov process. These photons travel close to the primary particle, 
forming a ``light pool'' whose characteristics can be calculated from 
elementary considerations and knowledge of the density profile of the 
atmosphere. Figure~1 illustrates the radial profile of the light pool for 
a variety of particle energies and nuclei. At 35~km altitude, attainable 
by a NASA balloon, the diameter of the lightpool increases from zero at 
the threshold energy to a maximum of about 15 meters for a fully 
relativistic particle incident 45 degrees from the zenith. The intensity 
has a central peak, a cusp at the edge of the pool, and a broad floor at 
intermediate radii. For an iron nucleus the floor is about 2000 photons 
per square meter in the sprectral band from 0.3 to 0.6 microns. With 
finite size light collectors the peak and cusp are softened, and for the 
24.3~cm radius light collectors employed by PDQ~BACH the floor 
corresponds to about 400 photons. Quantum efficiency of our system is 
about 0.1, so a minimum of 40 photoelectrons are detected in each 
collector. These arrive with a time spread of less than 0.1 ns and 
produce a single pulse when observed by a photomultiplier tube. The 
trigger resolving time of the instrument is determined by the approximate 
4~ns pulse width of the photomultipliers.

Cherenkov flashes from iron must be detected against the night sky 
background light. Fluctuations in response to background light can cause 
false event triggers reducing live time of the instrument, and 
sufficiently strong fluctuations may be confused with signals from cosmic 
ray iron nuclei. The PDQ BACH telemetry system takes 0.78 seconds to 
readout one event. To achieve livetime of 90\% we must limit the 
background trigger rate to 0.1 Hz. This means that the false trigger 
probability should be less than $4 \times 10^{-10}$ per resolving time. 
Operating two collectors in coincidence allows a fluctuation trigger rate 
$P = 2 \times 10^{-5}$ per resolving time in each collector.

The mean number of background photoelectrons in one sample is $N_B = 
\epsilon \phi_B A \Omega t$, where $\epsilon$ is the detection 
efficiency, $\phi_B$ is the background flux, $A$ is the collector area, 
$\Omega$ is the solid angle, and $t$ is the integration time.  
Pre-flight, we anticipated a dark sky background of $10^8 {\rm cm}^{-2} 
{\rm sr}^{-1} {\rm s}^{-1}$, which corresponds to $N_B \approx 4$ 
photoelectrons for BACH's design parameters.  For this $N_B$, a single 
collector threshold of 14 detected photoelectrons yields $P = 10^{-5}$. 
The equivalent photon count is shown in Figure~1 as the horizontal line 
labeled ``P5". For $N_B=8$ the corresponding threshold rises to 22 
photoelectrons. During flight, $N_B$ varied between 6 and 8 
photoelectrons, with occasional excursions as bright stars came into the 
field of view. At 40 photoelectrons, cosmic ray iron events lie well 
above the thresholds of 14-22 photoelectrons required from deadtime 
consideration. The fluctuations themselves are on the tail of a steep 
distribution, so an increase in threshold by a modest $\sim 15\%$ in 
post-flight analysis lowers the effective background rate to less than 
one per hour and allows for a clean separation of fluctuations and cosmic 
ray events. 

Lower $Z$ nuclei may be confused with iron if the cusp or central peak of 
the light pool falls on a collector. By separating the two coincident 
detectors by $\sim 3$ m one may ensure that the central peak 
cannot simultaneously overlay the two collectors, thus largely 
eliminating the background from low $Z$ nuclei. The cusps for CNO nuclei 
are not strong enough to generate triggers, but silicon events where the 
cusp falls on both collectors may mimic iron. As an illustration of the 
characteristics of the BACH instrument, Figure~2 shows the results of 
exposing a numerical model of the instrument to a simulated flux of 
cosmic rays. Panel (a) shows that silicon may provide some contamination 
for low intensity iron events, while panel (b) demonstrates that CNO 
nuclei do not. Details of the model are given in Section V.

We also considered other sources of background, such as Cherenkov 
radiation from 10 GeV electrons entering a collector ($\sim 50$ photons) 
and 100 GeV iron nuclei generating flourescent radiation directly in 
front of the collectors ($\sim 10$ photons), but neither has enough 
intensity to trigger the instrument. Numerous lower energy charged 
particles may pass through a detector generating spurious signals. By 
running two collectors in coincidence, the trigger rate from such 
processes is diminished. Offline analysis can distinguish between the few 
remaining particle events and true cosmic ray events.

In evaluating the viabilty of the BACH technique one must also consider 
details of the atmospheric density profile and factors that could alter 
the assumed straight line trajectory of the radiating nucleus. We assume 
a US standard atmosphere (NOAA 1976) profile with the density normalized 
to match the density of the atmosphere at the pressure altitude of the 
balloon. We have not modeled absorption due to ozone as the edge of that 
feature is obscured by absorption in the optical components of the 
instrument. We considered deflection by multiple Coulomb scattering and 
hadronic interactions (both $10^{-5}$ radians), as well as magnetic 
bending which may be a factor 5 larger depending on the track geometry. 
These effects are significantly smaller than $10^{-3}$~radians, the 
Cherenkov angle at altitude for fully relativistic particles, and so have 
not been included in our simulations of the experiment. Hadronic 
interactions eventually disintegrate the primary nucleus; however, most 
of these interactions are soft. Disintegration takes several interaction 
lengths and occurs mostly after the primary has passed the balloon. We 
estimate a 9\% exposure correction for this process, but do not model it 
in detail. 

\section{Instrumentation}

Our payload is illustrated schematically in Figure~3. Given the intrinsic 
100~psec duration of the Cherenkov light flash, the primary approach to 
skylight background is to reduce signal integration time. To this end, we 
employ Hamamatsu R4143 three inch diameter photomultipliers with a 
risetime of 2.4~nanoseconds at our operating point of 1200V. We also 
reduce background by limiting the acceptance aperture with light 
collectors (208.8~cm long, 24.3~cm radius aperture) based on concepts 
developed by Winston and colleagues (Welford and Winston 1989). These 
collectors have a nearly flat response within a well defined 7.5 degree 
half angle field of view. To reduce backgrounds, two collectors, 
separated on axis by 2.87~m, were operated in coincidence. The Winston 
collectors for PDQ~BACH were constructed using a precision-machined 
aluminum mandrel. Each collector consisted of two upper and two lower 
sections, molded individually on the mandrel from carbon fiber epoxy 
composite material. A reflective aluminum coating was evaporated onto the 
sections, which were then assembled. 

Photomultipliers attached to the collectors must operate in the presence 
of a comparatively high DC current from starlight, which must be 
monitored accurately while still maintaining satisfactory high frequency 
response. By operating with negative high voltage, the anode current can 
be directly measured; however, a photocathode (and, consequently, 
photomultiplier faceplate) maintained at negative high voltage while exposed to 
low atmospheric pressure poses an unacceptable risk of corona discharge. 
Our solution was to construct Pressurized Optical Detectors (PODs), as 
illustrated in Figure~4, to operate the photomultipliers at one 
atmosphere, and couple them to the collectors via light pipes of Bicron 
UV transparent acrylic. The light pipes extend out from the PODs to allow 
thermal isolation of the PODs and collectors, while providing electrical 
insulation as well. Surfaces of each light pipe are polished and kept from 
contacting the POD structure except for a 0.32~cm tapered flange that 
provides a pressure seal. Each light pipe is internally reflecting for 
almost all signal photon trajectories. The PODs also contain R329 
photomultipliers coupled to plastic scintillators that cover the detector 
photomultipliers and internal sections of the light pipes to serve as an 
anticoincidence shield against cosmic rays. Each photomultiplier receives 
high voltage from an individual power supply, adjustable in flight. 

Data were acquired with a Tektronics TDS640A digital oscilloscope which 
flew as part of the payload. This oscilliscope captures 2 gigasamples per 
second in four channels simultaneously. The four channels were used to 
record data from the two R4143 photomultipliers (denoted C1 and C2) and the two 
anticoincidence detectors (A1 and A2). Internal oscilloscope logic was 
used to detect coincident triggers where the signals in C1 and C2 both 
exceed thresholds. The thresholds were independently adjustable during 
flight. The anticoincidence channels were not part of the trigger, but 
were used in off line analysis. The TDS640A comes equipped with a GPIB 
interface, so we developed a microcontroller-based unit to interface it 
to the command and telemetry system. 

Crucial to the scientific return of a BACH payload is calibration of 
photomultiplier gain, pulse shape, and the efficiency of the light 
collectors. The full optical assemblies were calibrated for angular 
efficiency and alignment within the Ft. Sumner hanger (Seckel, et al 
1999). A 2.54~cm diameter diffuse light source was moved horizontally and 
vertically across the field of view at a distance of 7.5~m from the front 
of the collectors. The results of these tests were interpreted with the 
aid of ray tracing models for the collectors/PODs. The models include 
losses due to reflection from the collector (4\%), absorption by the 
aluminum coating (20\%), absorption in the light pipe (12\%), losses from 
the optical path (4\%), and yield an overall collector efficiency 
of 0.60 for a head on Cherenkov spectrum weighted by the photomultiplier 
quantum efficiency. Off axis, performance drops near the edge of the 
field of view in a calculable and testable manner. Minor adjustments were 
made to the individual collector models based on the test results. The 
adjusted models for the individual collectors were then used to determine 
instrument response to a light source at infinity.

Calibration of the photomultiplier response was complicated by low gain at the 
flight voltage of 1200 V and an inability to resolve pulses from single 
photoelectrons. We therefore followed a multistep process including 
determination of the absolute single photoelectron response at high 
voltage (2150 V), the relative gain curve from 2150 V down to 1200 V, and 
a comparison of charge and pulse height response at 1200 V. The mean 
pulse height per photoelectron for strong pulses at flight voltage was 
0.24 mV/pe in collector C1 and 0.19 mV/pe in collector C2. Single 
photoelectron pulse height distributions were determined at high voltage 
and extrapolated down to flight voltage using an 8 stage model for the 
R4143 photomultipliers.  

Ground level observations of cosmic rays provided a check on the absolute 
normalization of the collector and POD response. Muons passing through 
the light pipe serve as a rough check of the light pipe model and 
photomultiplier gain. Similarly, we are able to measure Cherenkov light 
from ground level muons and electrons passing through the collector 
volume on trajectories nearly parallel to the optical axis. We also 
performed ground level observations of Cherenkov light produced by cosmic 
ray air showers as a full system check of the payload in flight 
configuration.

\section{Observations}

PDQ~BACH was launched at sunset from Ft. Sumner, NM, on April 22,
1998 and cut down $\sim 6$ hours later, yielding a useful exposure
of 2.75 hours. Atmospheric pressure during the observation varied
from 8.05  to 8.5 mb, corresponding to a threshold energy for
iron of $2.2\times 10^{13}$ eV. The instrument collected 1793 triggers,
each consisting of 250 ns (500 samples) of waveform from four
channels: the two collectors and the two anticoincidence
detectors. Since the signals resulting from the Cherenkov light
are only a few nanoseconds wide, most of the waveform data are
available for monitoring the skylight background. Expecting only
a handful of true iron events, the trigger thresholds on the
detectors were intentionally set to acquire events at a rate of
approximately 0.1 Hz to allow detailed study of background
sources. 

Sample events from flight data are shown in Figure~5. Panel (a) contains 
one of the events identified as due to an iron nucleus. Note the 
identical arrival times of the similar sized pulses from the two 
collectors (C1 and C2) and the absence of signal from either 
anticoincidence (A1 and A2). Panel (b) shows a trigger with two moderate 
signals in C1 and C2. On the basis of pulse height, the event could be a 
silicon nucleus or iron incident at the edge of the field of view; 
however, there is a 2.5 ns offset in the two pulses, so this event is 
classified as a rare, high-amplitude skylight fluctuation. Panel (c) shows 
a typical event where a relativistic particle passes through one of the 
light pipes accompanied by a skylight fluctuation in the other channel. In 
addition to the signal in the anticoincidence detector, the event shows 
asymmetric response and a significant time lag between the C1 and C2 pulses. 
The time delay between collector (R4143) and anticoincidence (R329) signals 
results from the different transit times within the two types of 
photomultiplier. Panel (d) shows a rarer type of background event. The 
relative timing of the signals indicates that a particle passed through 
the light pipes and anticoincidence detectors in both of the PODs. 

Data from one, 1.75 hour, segment of the flight are shown in Figure~6. 
This represents the longest exposure taken with unchanged trigger 
conditions. Data obtained under other sets of trigger conditions were 
analyzed separately, and the final results were combined. Several events 
from the 1.75 hour segment have the exact characteristics that we expect 
from cosmic ray iron nuclei, namely strong coincident sharp signals well 
above threshold in both channels and no signal in either anticoincidence 
detector. Of the remainder, the majority have amplitudes just above 
threshold, and are most likely due to simultaneous fluctuations in the 
skylight background. As expected, the rate of such triggers is a strong 
function of the amplitude of the DC current observed in the waveforms 
outside the pulse region. We observed numerous asymmetric events where C1 
records a strong signal and C2 is just above threshold, resulting from 
charged particles passing through the light pipe in C1 accompanied by 
statistical fluctuations in C2. The rate of these light pipe events is 
also consistent with expectations. The relative lack of events where 
particle hits are evident in C2 coincident with a fluctuation in C1 
results from an asymmetry in trigger levels on the oscilloscope. This was 
intentional, to aid in controlling the trigger rate in a changing 
background. About 75\% of light pipe events are tagged by A1, consistent 
with our modeling for coverage by the anticoincidence detector. A few 
events have strong pulses in both channels but are clearly not caused by 
Cherenkov light, since both anticoincidence detectors also show signals. 
In all of these cases one signal leads the other by about 10 ns, 
consistent with a single charged particle passing through both 
light pipes. 

\section{Results}

After eliminating events with anticoindence hits, the data can be 
described as the sum of two distributions characterized by the time 
separation $dt$ of the two pulses. For true cosmic ray events, the 
signals arrive within a narrow coincidence window corresponding to the 
geometry of the collectors ($dt \sim 1$ ns). Skylight fluctuations have a 
broader distribution corresponding to the pulse width of the 
photomultipliers ($dt \sim 4$ ns). Events with strong signals are 
predominantly of the first type, whereas events with smaller signals are 
mostly in the second. We count cosmic ray events as those with signal 
greater than 8.5 mV in both channels and $dt<0.75$ ns. There are 9 such 
events. Some of these events could be skylight events which just happen 
to have $dt<0.75$. By comparing to the clean skylight sample, i.e. events 
with $dt>0.75$ ns, we estimate a sklylight background count of 0.75 
events. We also estimate that inefficiency of the anticoincidence 
detectors has resulted in false acceptance of 1 event, leaving a total of 
7.25 cosmic ray events. 

To convert this number to an iron flux, we compare to the mathematical 
model of the instrument exposed to a flux of cosmic ray nuclei of assumed 
composition and spectrum, as illustrated in Figure~2. The incident flux 
is taken to be isotropic over the field of view of the collectors with 
impact parameters that allow for all positions and orientations of the 
payload. Knowledge of the light pool, impact parameters, and 
POD/collector efficiency are combined to yield an expected number of 
photoelectrons, which in turn is the basis for drawing a signal amplitude 
from a distribution based on Poisson statistics, the photomultiplier single 
photoelectron response and the modeled pulseheight distributions for the 
photomultipliers. Contribution of background light to the signal is 
modeled by overlaying livetime weighted randomly chosen sections of 
waveform from the pre-trigger flight data. Pairs of pulseheights thus 
determined are displayed in the scatterplots of Figure~2. Applying the 
8.5 mV cut to the simulation yields a count which is directly comparable 
to the 7.25 counts from analysis of the flight data.

Normalizing the exposure to the simulation, yields a flux of 
$\phi_{Fe}(5.05^{+4.5}_{-1.6}~\times~10^4{\rm~GeV}) = 
677^{+192}_{-139}~~E^{-2.5}$ m$^{-2}$ sr$^{-1}$ s$^{-1}$ GeV$^{1.5}$. We 
use Poisson statistics to estimate the statistical error. The energy 
value (range) is the peak (width at half maximum) of the distribution 
obtained by convoluting the instrument response with an assumed 
$E^{-2.5}$ spectrum.  The result is reasonably independent of the 8.5 mV 
analysis threshold, but depends on the assumed iron spectral index. With 
a small number of events and uncertain calibration we cannot determine 
the spectrum. Our result also depends on the composition, as silicon 
nuclei with particular impact parameters can imitate the signals from 
iron. For the 8.5~mV threshold analysis, the result includes a 6$\%$ 
correction to account for mis-identified silicon. Corrections were also 
made for hadronic interactions of iron in the atmosphere (estimated 9 
$\%$), and for contamination of the iron flux by nuclei with Z $=$ 17-25 
(estimated 10$\%$). 

Figure~7 compares our new determination of the iron flux with earlier 
measurements.  Our results are consistent with both the CRN and JACEE 
determinations. All of the recent data are above the flux reported by 
Sood (1983). We have not reanalyzed Sood's experiment, but have attempted 
to improve on his pioneering effort in several ways to make our 
instrument easier to model and the interpretation of the data more 
certain.

\acknowledgments We thank A. McDermott, L. Piccirillo, E. Rode, C. 
Scharmberg, L. Shulman and J. Poirier for technical assistance. The 
collectors were produced in collaboration with the University of Bremen, 
the Astronomical Institute in Hamburg, and the private firm ``Surf 
Affairs'', located near Kiel. The thank the National Scientific Balloon 
Facility for the balloon flight and the NSBF staff for preflight 
assistance. The work was supported in part by NASA grant NAG5-5063 and 
NAG5-5221. 




\newpage

\figcaption[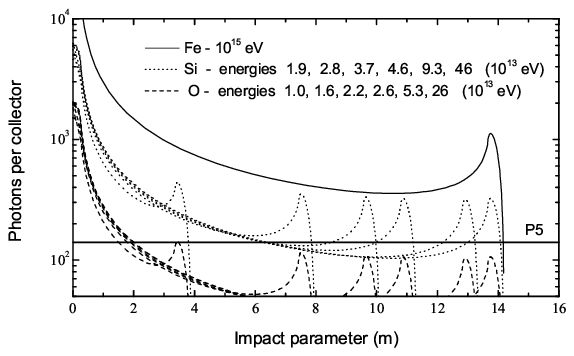]{
Yield of Cherenkov photons as a function of distance perpendicular 
to the primary trajectory (impact parameter) for primaries incident at a 
zenith angle of 45 deg, a balloon altitude of 35,000 m, and a collector 
radius of 24.3~cm. The horizontal line labeled ``P5" is the signal 
threshold required to reduce the rate of fluctuation triggers (for two 
collectors in coincidence) to less than 0.1 Hz under dark sky conditions. 
Cherenkov flashes from silicon nuclei may be confused with iron for some 
impact parameters, but an instrument with two collectors separated by 3 
meters will never be triggered by oxygen nuclei. 
\label{fig1}}

\figcaption[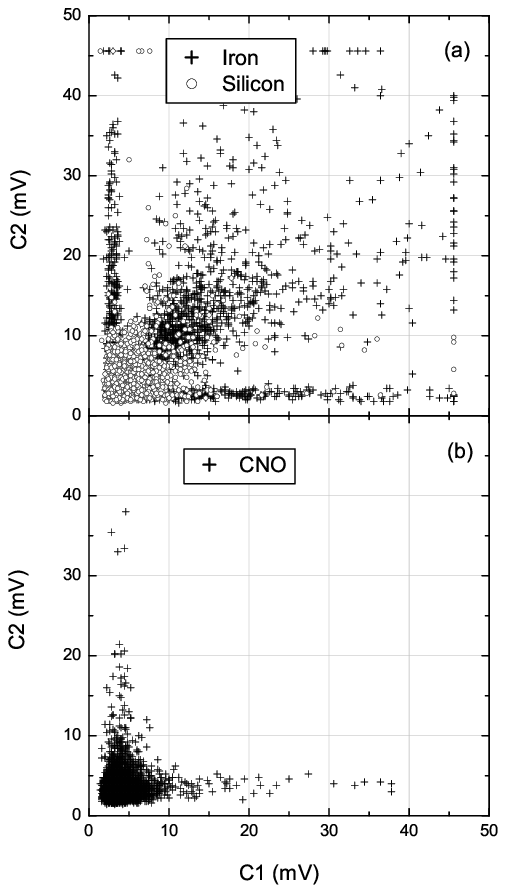]{
Simulated long duration balloon exposure (6 collector pairs, 6 hrs per 
night, 15 day flight) of the BACH payload to cosmic ray Fe, Si, C, N, and 
O with differential spectra proportional to E$^{-2.5}$ and constant 
composition extrapolated from lower energy data. Silicon events in panel 
(a) obscure some underlying iron events. Carbon, nitrogen and oxygen are 
shown separately in panel (b) for visibility. Results are expressed as 
simulated pulse heights (in mV) from two collectors (C1 and C2) given the 
instrument parameters and atmosphere of the PDQ-BACH flight.
\label{fig2}}

\figcaption[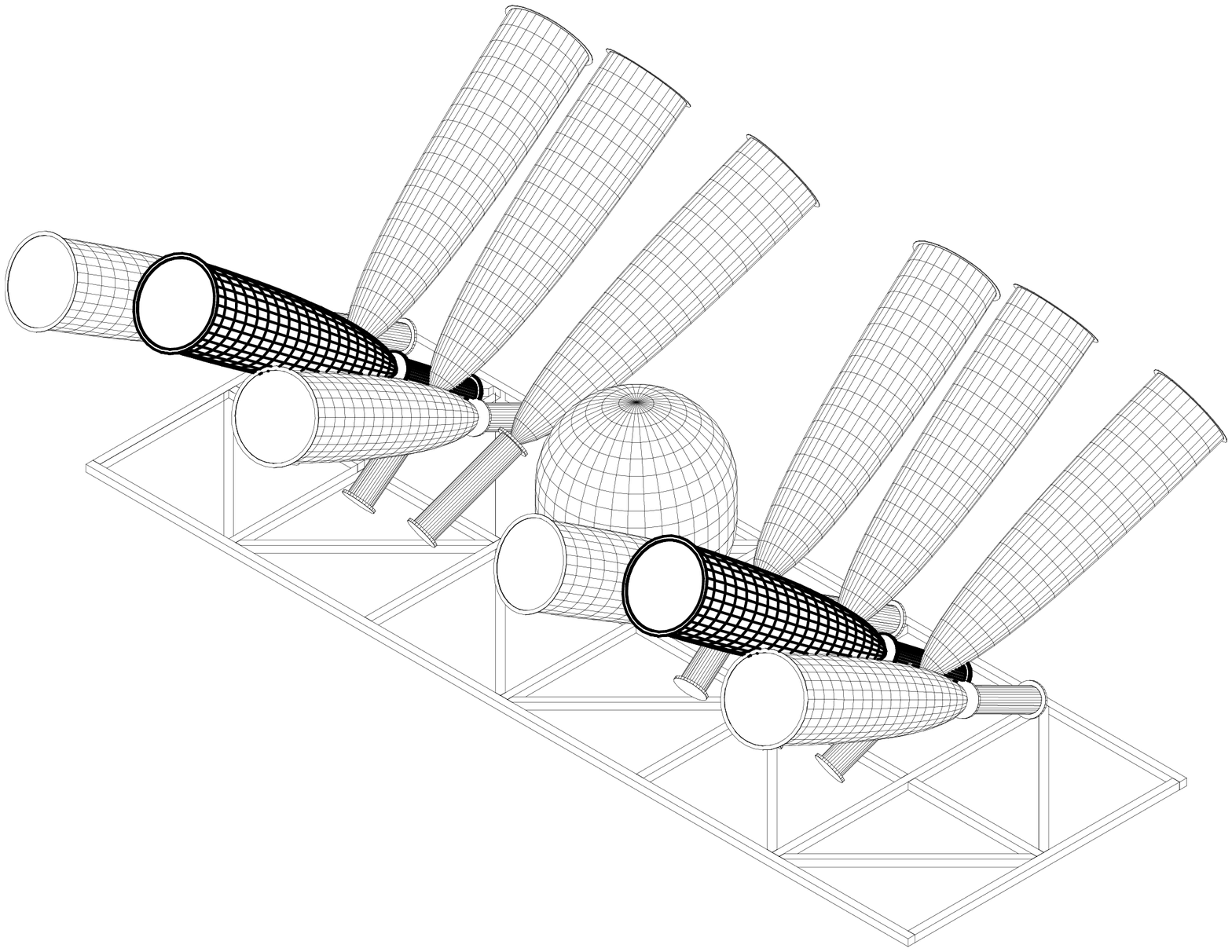]{
Balloon Air CHerenkov (BACH) long duration balloon payload design with 
250 m$^2$-sr geometry factor. PDQ~BACH consisted of the two highlighted 
detector assemblies and associated electronics. Centerline separation of 
the collectors was 2.87 meters for PDQ~BACH. 
\label{fig3}}

\figcaption[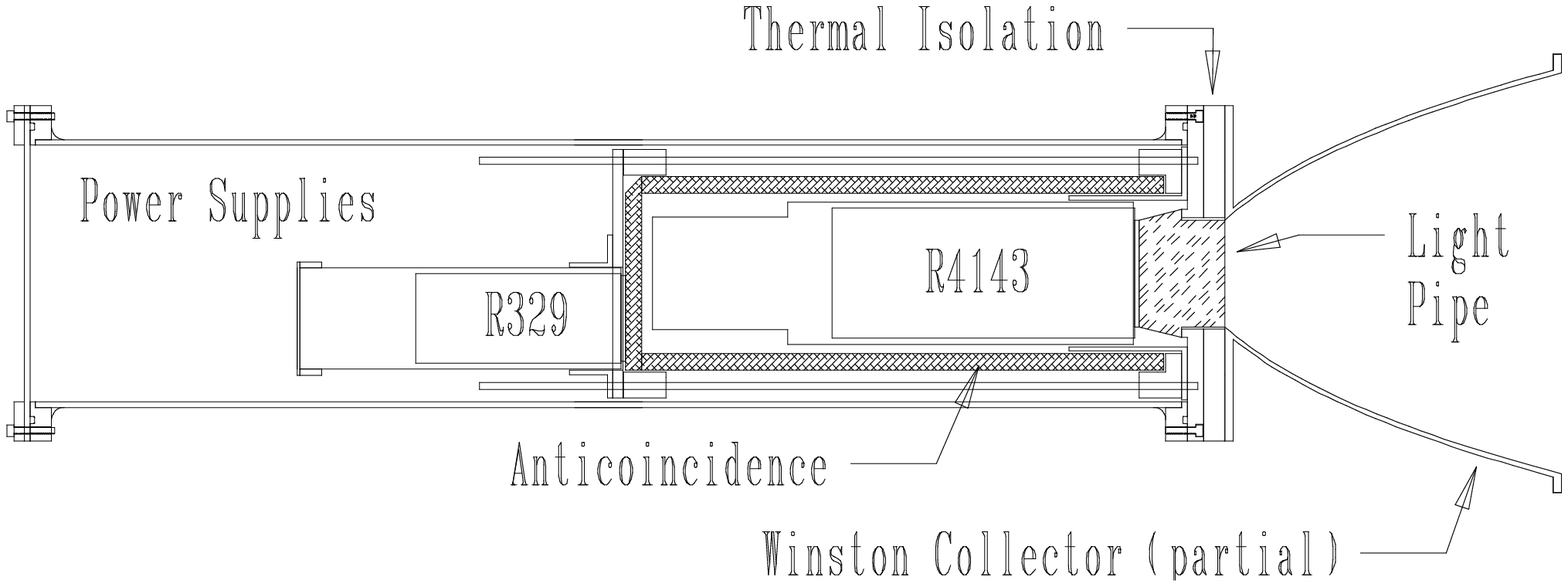]{
Pressurized Optical Detectors (PODs) utilized by PDQ BACH. 
\label{fig4}}

\figcaption[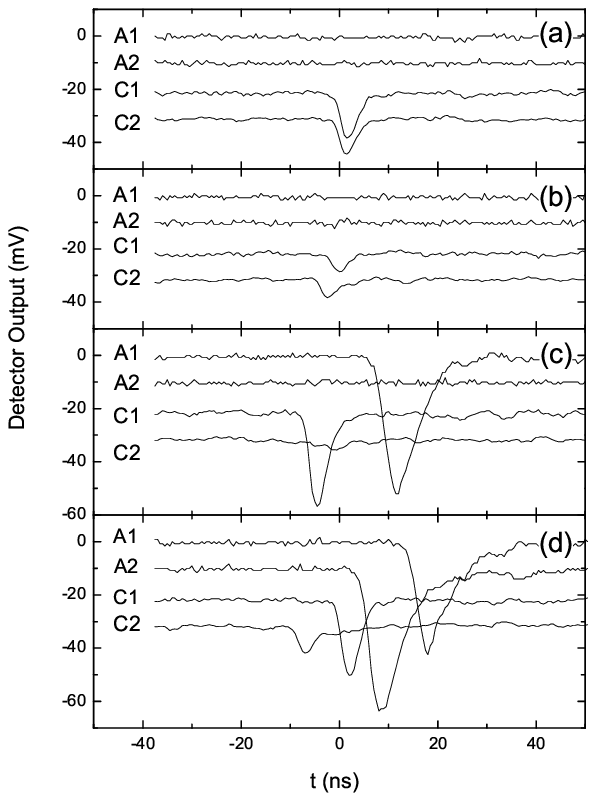]{
Oscilloscope traces resulting from four triggers during the PDQ~BACH 
flight. Traces are offset in units of 10 mV. Strong, coincident pulses in 
C1 and C2, along with an absence of signal in anticoincidence detectors 
A1 and A2, identify the event in panel (a) as resulting from an iron 
nucleus. The timing offset of 2.5 ns tags the event in panel (b) as  
a skylight fluctuation. Panel (c) illustrates a background event where a 
charged particle is detected in C1 and its anti A1, while a skylight 
fluctuation occurs in C2. In panel (d), the presence of signals in 
A1 and A2, and the 10 ns offset between the C1 and C2 pulses indicates a 
charged particle passing through both light pipes.
\label{fig5}}

\figcaption[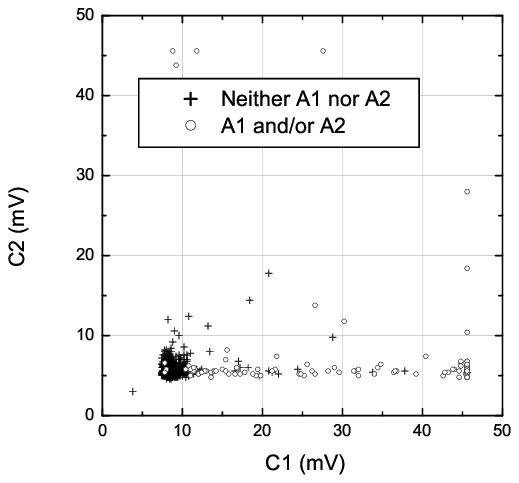]{
Flight data from PDQ~BACH during the first 1.75 hrs of operation. These 
data may be compared to the simulation shown in Figure 2. The flight data 
include skylight and charged particle backgrounds, whereas the 
simulation contains only cosmic ray air Cherenkov events.
\label{fig6}}

\figcaption[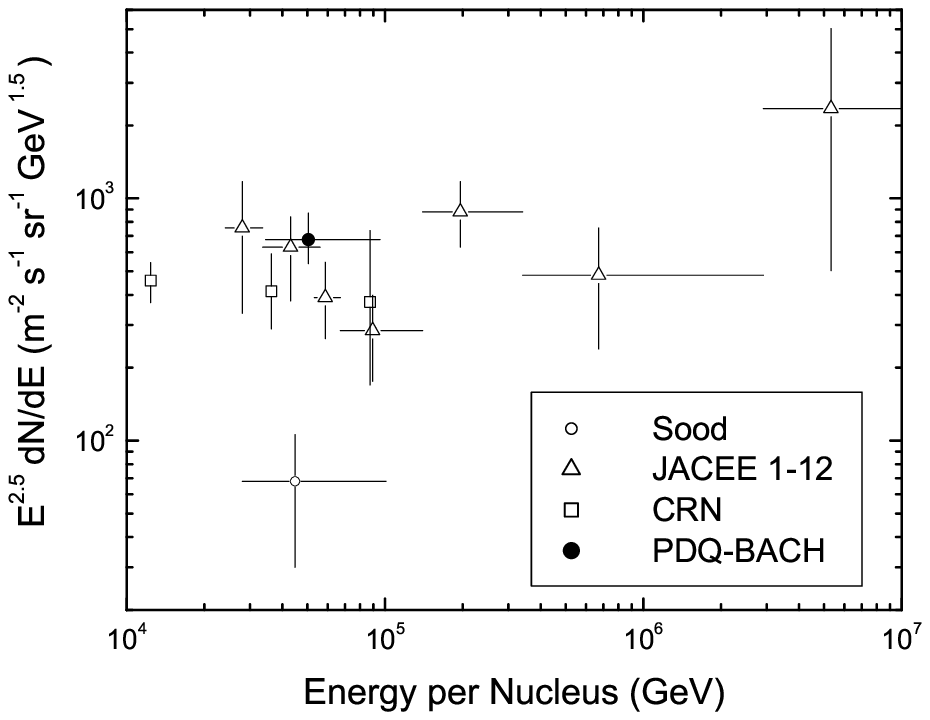]{
Determinations of the cosmic ray iron flux by other investigators
compared to results from PDQ~BACH. Sood (1983) pioneered the BACH
technique. The JACEE determination (Asakimori et al. 1995) is a
compilation of data from 12 flights. CRN (Swordy et al. 1993) was
conducted on the Space Shuttle. 
\label{fig7}}


\begin{figure}
\epsscale{1}
\plotone{fig1}
\centerline{Figure~1}
\end{figure}

\begin{figure}
\epsscale{.6}
\plotone{fig2}
\centerline{Figure~2}
\end{figure}

\begin{figure}
\epsscale{1}
\plotone{fig3}
\centerline{Figure~3}
\end{figure}

\begin{figure}
\epsscale{1}
\plotone{fig4}
\centerline{Figure~4}
\end{figure}

\begin{figure}
\epsscale{1}
\plotone{fig5}
\centerline{Figure~5}
\end{figure}

\begin{figure}
\epsscale{1}
\plotone{fig6}
\centerline{Figure~6}
\end{figure}

\begin{figure}
\epsscale{1}
\plotone{fig7}
\centerline{Figure~7}
\end{figure}

\end{document}